\documentclass{aa}

\usepackage{graphicx}
\usepackage{txfonts}
\def\lsim{\mathrel{\rlap{\lower4pt\hbox{\hskip1pt$\sim$}}
    \raise1pt\hbox{$<$}}}
\def\gsim{\mathrel{\rlap{\lower4pt\hbox{\hskip1pt$\sim$}}
    \raise1pt\hbox{$>$}}}
\begin{document}

   \title{Constraints on the generalized Chaplygin gas model including gamma-ray bursts via a Markov Chain Monte
Carlo approach}
   \author{Nan Liang \inst{1}, Lixin Xu \inst{2,3}          \and    Zong-Hong Zhu \inst{1}    }
   \institute{Department of Astronomy, Beijing Normal   University, Beijing 100875, China  \and School of Physics and
Optoelectronic Technology, Dalian University of Technology, Dalian,
Liaoning 116024, P. R. China \and College of Advanced Science \&
Technology, Dalian University of Technology, Dalian, 116024, P. R. China\\
   \email{liangn@bnu.edu.cn, lxxu@dlut.edu.cn,  zhuzh@bnu.edu.cn}}

\abstract {} {We investigate observational constraints on the
generalized Chaplygin gas (GCG) model  including the gamma-ray
bursts (GRBs) at high redshift obtained directly from the Union2
type Ia supernovae (SNe Ia) set.} { By using the Markov Chain Monte
Carlo method, we constrain the GCG model with the
cosmology-independent GRBs, as well as the Union2 set, the cosmic
microwave background (CMB) observation from the Wilkinson Microwave
Anisotropy Probe (WMAP7) result, and the baryonic acoustic
oscillation (BAO) observation from the spectroscopic Sloan Digital
Sky Survey (SDSS) data release 7 (DR7) galaxy sample.} {The best-fit
values of the GCG model parameters are
$A_S$=$0.7475_{-0.0539}^{+0.0556}(1\sigma)_{-0.0816}^{+0.0794}(2\sigma)$,
$\alpha$=$-0.0256_{-0.1326}^{+0.1760}(1\sigma)_{-0.1907}^{+0.2730}(2\sigma)$,
and the effective matter density $\Omega_{\rm
m}=0.2629_{-0.0153}^{+0.0155}(1\sigma)_{-0.0223}^{+0.0236}(2\sigma)$,
which are more stringent than previous results for constraining GCG
model parameters.} {}


\authorrunning{N. Liang, L. Xu and Z.-H. Zhu}
\titlerunning{Constraints on GCG model including GRBs}

\maketitle

\section{Introduction}

The original Chaplygin gas (CG, Kamenshchik et al. 2001) and
generalized Chaplygin gas models (GCG, Bento et al. 2002) have been
proposed as possible explanations of the acceleration of the current
universe, with the equation of state as follows
\begin{equation}
p_{\rm GCG} = -\frac{A}{\rho_{\rm GCG}^{\alpha}},
\end{equation}
where $A$ and $\alpha$ are two parameters to be determined. For the
case $\alpha = 1$, it corresponds to the original Chaplygin gas
(Kamenshchik et al. 2001); if $\alpha = 0$, it acts as the
cosmological constant ($\Lambda$). Considering the relativistic
energy conservation equation in the framework of
Friedmann-Robertson-Walker (FRW) metric, we can obtain
\begin{equation} \rho_{{\rm GCG}} =
 \rho_{{\rm GCG},0} \left[A_s +
 (1-A_s)a^{-3(1+\alpha)}\right]^{1\over(1+\alpha)},
\end{equation}
where  $A_s \equiv A/\rho_{{\rm GCG},0}^{1 + \alpha}$, $\rho_{{\rm
GCG},0}$ is the energy densities of the GCG at present,  and the
scale factor is related to the redshift by $a = 1/(1+z)$. From
Equation (2), the striking property of the GCG can be found that the
energy density behaves as dust like matter at early times; while it
behaves like a cosmological constant at late times. Therefore, the
GCG model can be regarded as a derivative of the unified dark
matter/energy (UDME) scenario (Bento et al. 2004). Until now, the
GCG model has been constrained using many different types of
observational data, such as Type Ia supernovae  (SNe Ia)
(Fabris et al. 2002;    Makler et al. 2003a;   
Colistete et al. 2003; Silva and Bertolami 2003;   Cunha et al.
2004; Bertolami et al. 2004;  Bento et al. 2006; 
Wu and  Yu 2007a), cosmic microwave background (CMB) anisotropy
(Bento et al. 2003a, 2003b; Bean and  Dore 2003; Amendola et al.
2003), the angular size of the compact radio sources (Zhu 2004), the
X-ray gas mass fraction of clusters (Cunha et al. 2004; Makler et
al. 2003b), the Hubble parameter versus redshift data (Wu and  Yu
2007b), large-scale structure (Bili\'c et al. 2002; Multam\"aki et
al. 2004), gravitational lensing surveys (Dev et al. 2003, 2004;
Chen 2003a, 2003b), age measurements of high-$z$ objects (Alcaniz et
al. 2003) and  lookback time of galaxy clusters (Li, Wu and Yu
2009); as well as various combinations of data (Wu and Yu 2007c;
Davis et al. 2007; Li, Li and Zhang  2010; Xu and  Lu 2010).

Gamma-ray bursts (GRBs) have been proposed as distance indicators
and regarded as a complementary cosmological probe of the universe
at high redshift (Schaefer 2003; Dai et al. 2004; Ghirlanda et al.
2004; Firmani et al. 2005, 2006; Liang and Zhang 2005; Ghirlanda et
al. 2006; Schaefer 2007; Wang et al. 2007; Wright 2007; Amati 2008;
Basilakos and Perivolaropoulos 2008; Mosquera Cuesta et al. 2008a,
2008b; Daly et al. 2008). Owing to the lack of a low-redshift
sample, the empirical luminosity relations of GRBs had been usually
calibrated by assuming a certain cosmological model with particular
model parameters. Liang et al. (2008) presented a completely
cosmology-independent method to calibrate GRB luminosity relations
with the luminosity distances of GRBs at low redshift interpolated
directly from SNe Ia or  by other similar  approaches (Liang and
Zhang 2008; Kodama et al. 2008; Cardone et al. 2009; Gao et al.
2010; Capozziello and Izzo 2010). Following the
cosmology-independent calibration method, the derived GRB data at
high redshift can be used to constrain cosmological models by using
the standard Hubble diagram method (Capozziello and Izzo 2008; Izzo
et al. 2009; Wei and Zhang 2009; Wei 2009;  Qi et al. 2009; Wang et
al. 2009a, 2009b; Liang, Wu and Zhang 2010; Wang and Liang 2010;
Liang, Wu and Zhu 2010; Wei 2010a, 2010b; Liang and Zhu 2010;
Demianski et al. 2010). Bertolami and Silva (2006) first studied the
GCG model by considering the use of GRBs at $1.5 < z < 5$ calibrated
with the bursts at $z<1.5$ as distance markers. The joint analysis
with the GCG model of the cosmology-independent GRB data set
obtained in Liang et al. (2008) can be found in Wang et al. (2009a)
and Freitasa et al. (2010).

Liang, Wu and Zhu (2010) calibrate GRBs data at high redshift
directly from the Union2 compilation of 557 SNe Ia data set
(Amanullah et al. 2010) and constrain the Cardassian models by
combining the updated GRB data with the joint observations. In this
paper,  we use the Markov Chain Monte Carlo (MCMC) technique to
constrain the GCG model from the latest observational data including
the updated distance moduli of the GRBs at high redshift obtained
directly from the Union2 set. We combine the GRB data with both the
joint observations, such as the Union2 set, the CMB observation from
the Wilkinson Microwave Anisotropy Probe (WMAP7; Komatsu et al.
2010) result, and the baryonic acoustic oscillation (BAO)
observation from the spectroscopic Sloan Digital Sky Survey (SDSS)
data release 7 (DR7) galaxy sample (Percival et al. 2010).

\section{Observational Data Analysis}
The Union2 compilation  consists of  data for 557 SNe Ia (Amanullah
et al. 2010), and we use the  69 GRBs data compiled by Schaefer
(2007). Following Liang, Wu and Zhu (2010), we use the updated
distance moduli of the 42 GRBs at $z>1.4$, which calibrated with the
sample at $z\le1.4$ by using the linear interpolation method from
the Union2 set.  For more details for the calculations for GRBs, we
refer to Liang et al. (2008) and Liang, Wu and Zhang (2010).
Constraints from SNe Ia and GRB data can be obtained by fitting the
distance moduli $\mu(z)$. A distance modulus can be calculated as
\begin{eqnarray}\label{mu}
\mu=5\log \frac{d_L}{\mathrm{Mpc}} + 25=5\log_{10}D_L-\mu_0,
\end{eqnarray}
where $\mu_0=5\log_{10}[H_0/(100{\rm km/s/Mpc})]+42.38$, and the
luminosity distance $D_L$ can be calculated  using
\begin{eqnarray}
D_L\equiv
H_0d_L=(1+z)\Omega_\mathrm{k}^{-1/2}\mathrm{sinn}\bigg[\Omega_\mathrm{k}^{1/2}\int_0^z\frac{dz'}{E(z')}\bigg],
\end{eqnarray}
where  $\rm{sinn}$$(x)$ is $\rm sinh$ for $\Omega _{\rm k}>0$, $\rm
sin$ for $\Omega _{\rm k}<0$, and $x$ for $\Omega _{\rm k}=0$, and
$E(z)=H/H_0$, which is determined by the choice of the specific
cosmological model. The $\chi^2$ value of the observed distance
moduli can be calculated by
\begin{eqnarray}
\chi^2_{\mu}=\sum_{i=1}^{N}\frac{[\mu_{\mathrm{obs}}(z_i)-\mu(z_i)]^2}
{\sigma_{\mu,i}^2},
\end{eqnarray}
where $\mu _{\mathrm obs}(z_i)$ are the observed distance modulus
for the SNe Ia and/or GRBs at redshift $z_i$ with its error
$\sigma_{\mu_{\mathrm i}}$; $\mu(z_i)$ are the theoretical value of
the distance modulus from cosmological models. Following the
effective approach (Nesseris and Perivolaropoulos 2005), we
marginalize the nuisance parameter $\mu_0$ by minimizing
${\hat\chi}^2_{\mu}=C- {B^2}/{A}$,
where $A=\sum{1}/{\sigma_{\mu_{i}}^2}$,
$B=\sum{[\mu_{\mathrm{obs}}(z_i)-5\log_{10}D_L]}/{\sigma_{\mu_{i}}^2}$,
and
$C=\sum{[\mu_{\mathrm{obs}}(z_i)-5\log_{10}D_L]^2}/{\sigma_{\mu_{i}}^2}$.

For the CMB observation, we use the data set including  the acoustic
scale ($l_a$), the shift parameter ($R$), and the redshift of
recombination ($z_{\ast}$), which provide an efficient summary of
CMB data as far as cosmological constraints go. The acoustic scale
can be expressed as
\begin{equation}
l_a=\pi\frac{\Omega_\mathrm{k}^{-1/2}sinn[\Omega_\mathrm{k}^{1/2}\int_0^{z_{\ast}}\frac{dz}{E(z)}]/H_0}{r_s(z_{\ast})},
\end{equation}
where $r_s(z_{\ast})
={H_0}^{-1}\int_{z_{\ast}}^{\infty}c_s(z)/E(z)dz$ is the comoving
sound horizon at photo-decoupling epoch. The shift parameters can be
expressed as
\begin{equation}
R=\Omega_{\mathrm{M0}}^{1/2}\Omega_\mathrm{k}^{-1/2}sinn\bigg[\Omega_\mathrm{k}^{1/2}\int_0^{z_{\ast}}\frac{dz}{E(z)}\bigg].
\end{equation} The redshift of
recombination  can be given by (Hu and Sugiyama 1996)
\begin{equation}
z_{\ast}=1048[1+0.00124(\Omega_bh^2)^{-0.738}(1+g_{1}(\Omega_{\mathrm{M0}}h^2)^{g_2})],
\end{equation}
where
$g_1=0.0783(\Omega_bh^2)^{-0.238}(1+39.5(\Omega_bh^2)^{-0.763})^{-1}$
and $g_2=0.560(1+21.1(\Omega_bh^2)^{1.81})^{-1}$. From the WMAP7
measurement, the best-fit values of the data set ($l_a$, $R$,
$z_{\ast}$) are (Komatsu et al. 2010)
\begin{eqnarray}
\hspace{-.5cm}\bar{\textbf{P}}_{\rm{CMB}} &=& \left(\begin{array}{c}
{\bar l_a} \\
{\bar R}\\
{\bar z_{\ast}}\end{array}
  \right)=
  \left(\begin{array}{c}
302.09 \pm 0.76\\
1.725\pm 0.018\\
1091.3 \pm 0.91 \end{array}
  \right).
 \end{eqnarray}
The $\chi^2$ value of the CMB observation can be expressed as
(Komatsu et al. 2010)
\begin{eqnarray}
\chi^2_{\mathrm{CMB}}=\Delta
\textbf{P}_{\mathrm{CMB}}^\mathrm{T}{\bf
C_{\mathrm{CMB}}}^{-1}\Delta\textbf{P}_{\mathrm{CMB}},
\end{eqnarray}
where $\Delta\bf{P_{\mathrm{CMB}}} =
\bf{P_{\mathrm{CMB}}}-\bf{\bar{P}_{\mathrm{CMB}}}$, and the
corresponding inverse  covariance matrix is
\begin{eqnarray}
\hspace{-.5cm} {\bf C_{\mathrm{CMB}}}^{-1}=\left(
\begin{array}{ccc}
2.305 &29.698 &-1.333\\
29.698 &6825.270 &-113.180\\
-1.333 &-113.180 &3.414
\end{array}
\right).
\end{eqnarray}

For the BAO observation, we use the measurement of the BAO distance
ratio ($d_z$) at $z=0.2$ and $z=0.35$ (Percival et al. 2010), which
can be expressed as
\begin{equation}
d_z=\frac{r_s(z_d)}{D_V(z_{\mathrm{BAO}})},
\end{equation}
where the distance scale $D_V$ is given by (Eisenstein et al. 2005)
\begin{equation} D_V(z_{\mathrm{BAO}})=\frac{1}{H_0}\big
[\frac{z_{\mathrm{BAO}}}{E(z_{\mathrm{BAO}})}\big(\int_0^{z_{\mathrm{BAO}}}\frac{dz}{E(z)}\big
)^2\big]^{1/3}~,
\end{equation}
and $r_s(z_d)$ is the comoving sound horizon at  the drag epoch at
which baryons were released from photons, $z_d$ can be given by
(Eisenstein \& Hu 1998)
\begin{equation}
z_{d}=\frac{1291(\Omega_{\mathrm{M0}}h^2)^{0.251}}{[1+0.659(\Omega_\mathrm{M0}h^2)^{0.828}]}[(1+b_{1}(\Omega_{b}h^2)^{b_2})],
\end{equation}
where
$b_1=0.313(\Omega_{\mathrm{M0}}h^2)^{-0.419}[1+0.607(\Omega_{\mathrm{M0}}h^2)^{0.674}]^{-1}$
and $b_2=0.238(\Omega_{\mathrm{M0}}h^2)^{0.223}$. From SDSS data
release 7 (DR7) galaxy sample, the best-fit values of the data set
($d_{0.2}$, $d_{0.35}$)   are (Percival et al. 2010)
\begin{eqnarray}
\hspace{-.5cm}\bar{\bf{P}}_{\rm{BAO}} &=& \left(\begin{array}{c}
{\bar d_{0.2}} \\
{\bar d_{0.35}}\\
\end{array}
  \right)=
  \left(\begin{array}{c}
0.1905\pm0.0061\\
0.1097\pm0.0036\\
\end{array}
  \right).
 \end{eqnarray}
The $\chi^2$ value of the BAO observation from SDSS DR7 can be
expressed as (Percival et al. 2010)
\begin{eqnarray}
\chi^2_{\mathrm{BAO}}=\Delta
\textbf{P}_{\mathrm{BAO}}^\mathrm{T}{\bf
C_{\mathrm{BAO}}}^{-1}\Delta\textbf{P}_{\mathrm{BAO}},
\end{eqnarray}
where  the corresponding inverse  covariance matrix is
\begin{eqnarray}
\hspace{-.5cm} {\bf C_{\mathrm{BAO}}}^{-1}=\left(
\begin{array}{ccc}
30124& -17227\\
-17227& 86977\\
\end{array}
\right).
\end{eqnarray}

\section{CONSTRAINTS ON THE GCG MODEL VIA MCMC METHOD}
We consider a flat universe filled with  the GCG component and the
baryon matter component. From the
Friedmann equation %
$H^2 = (8\pi G/3) (\rho_{\rm b}+\rho_{\rm GCG})$,
we find that
    \begin{eqnarray}
\label{eq:newE} E^2(z; A_s, \alpha) \equiv\frac{H^2}{H_0^2} & = &
   \Omega_{\rm b}(1+z)^3+(1-\Omega_{\rm b})
   \nonumber\\
   & &\times\left[A_s
   +(1-A_s)(1+z)^{3(1+\alpha)}\right]^{1\over(1+\alpha)},
\end{eqnarray}
where $\Omega_{\rm b}$ represents the fractional contribution of
baryon matter. The effective matter density in the GCG model can be
given by (Bento et al. 2004; Wu and  Yu 2007c)
\begin{eqnarray}
\Omega_{\rm m}=\Omega_{\rm b}+(1-\Omega_{\rm
b})(1-A_s)^{1\over(1+\alpha)}.
\end{eqnarray}

To combine GRB data with the SNe Ia data and constrain cosmological
models, we follow the simple method of avoiding any correlation
between the SNe Ia data and the GRB data: the 40 SNe points used in
the interpolating procedure are excluded  from the Union2 SNe Ia
sample used to derive the joint constraints (Liang, Wu and Zhang
2010; Liang, Wu and Zhu 2010). The 42 GRBs and the reduced 517 SNe
Ia, CMB, BAO are all effectively independent, therefore we can
combine the results by simply multiplying the likelihood functions.
The total $\chi^2$ with the SNe + GRBs + CMB + BAO  dataset is
\begin{eqnarray}
\chi^2={\hat\chi}^2_\mathrm{\{SNe,GRBs\}}+\chi^2_\mathrm{CMB}+\chi^2_\mathrm{BAO}.
\end{eqnarray}

We perform a global fitting to determine the cosmological parameters
using the Markov Chain Monte Carlo (MCMC) method. In adopting the
MCMC approach, we generate  using Monte Carlo methods a chain of
sample points distributed in the parameter space according to the
posterior probability, using the Metropolis-Hastings algorithm with
uniform prior probability distribution. In the parameter space
formed by the constraint cosmological parameters, a random set of
initial values of the model parameters is chosen to calculate the
$\chi^2$ or the likelihood. Whether the set of parameters can be
accepted as an effective Markov chain or not is determined by the
Metropolis-Hastings algorithm. The accepted set not only forms a
Markov chain, but also provides a starting point for the next
process. We then repeat this process until the established
convergence accuracy can be satisfied.  The convergence is tested by
checking the so-called worst e-values [the
variance(mean)/mean(variance) of 1/2 chains] $R-1 < 0.005$.

Our MCMC code is based on the publicly available {\bf CosmoMC}
package (Lewis \& Bridle 2002), and we generated eight chains after
setting $R-1 = 0.001$ to guarantee the accuracy of this work. We
show the 1-D probability distribution of each parameter in the MCMC
method ($\Omega_{\rm{b}}h^2$, $A_S$, $\alpha$, $\Omega_{\Lambda}$,
Age/Gyr, $\Omega_{\rm{m}}$, $H_0$) and 2-D plots for parameters
between each other for the GCG model with SNe + GRBs + CMB +  BAO in
figure 1 (Age/Gyr is the cosmic age, in units of Gyr). The best-fit
values of the GCG model parameters with the joint observational data
are
$A_S$=$0.7475_{-0.0539}^{+0.0556}(1\sigma)_{-0.0816}^{+0.0794}(2\sigma)$,
$\alpha$=$-0.0256_{-0.1326}^{+0.1760}(1\sigma)_{-0.1907}^{+0.2730}(2\sigma)$,
and the effective matter density $\Omega_{\rm
m}=0.2629_{-0.0153}^{+0.0155}(1\sigma)_{-0.0223}^{+0.0236}(2\sigma)$.
For comparison, fitting results from the joint data of 557 SNe Ia,
the CMB and BAO without GRBs and 42 GRBs, the CMB and BAO without
SNe Ia are given in Figs 2 and 3. We present the best-fit values of
each parameter with the 1-$\sigma$ and 2-$\sigma$ uncertainties, as
well as $\chi_{\rm min}^2$, in Table 1.

From Figs. 1-3 and Table 1, it is shown that the cosmological
constant ($\alpha = 0$) is allowed at the 1-$\sigma$ confidence
level, and  the original Chaplygin gas model ($\alpha = 1$) is ruled
out at $95.4\%$ confidence level, which are both consistent with
that obtained in Wu and Yu (2007c), and Li, Wu and Yu (2009). We can
find that GRBs can provide strong constraints when combined with CMB
and BAO data without SNe Ia, which has been also noted by Liang, Wu
and Zhu (2010), Liang and Zhu (2010), and Gao et al. (2010). In
additoin, the constraining results in this work with the joint
observational data including GRBs are more stringent than previous
results for constraining GCG model parameters with GRBs and/or other
combined observations (e.g. Wang et al. 2009a, 2009b; Freitasa et
al. 2010; Davis et al. 2007; Wu and Yu 2007c; Li, Wu and Yu 2009;
Li, Li and Zhang 2010; Xu and Lu 2010).

\begin{table*}
 \begin{center}
 \begin{tabular}{|c|c|c|c|} \hline\hline
 & \multicolumn{3}{c|}{The GCG Model}  \\
 \cline{2-4}                 &     SNe+GRBs+CMB+BAO                  &     SNe+CMB+BAO                       &     GRBs+CMB+BAO          \\ \hline
$\Omega_{\rm{b}}h^2$  \ \ & \ \ $0.0222_{-0.0007}^{+0.0008}(1\sigma)_{-0.0012}^{+0.0009}(2\sigma)$ \ \  & \ \ $0.0222_{-0.0007}^{+0.0008}(1\sigma)_{-0.0009}^{+0.0012}(2\sigma)$ \ \ & \ \ $0.0222_{-0.0007}^{+0.0008}(1\sigma)_{-0.0009}^{+0.0012}(2\sigma)$\ \ \\
$A_S$                 \ \ & \ \ $0.7475_{-0.0539}^{+0.0556}(1\sigma)_{-0.0816}^{+0.0794}(2\sigma)$ \ \  & \ \ $0.7668_{-0.0510}^{+0.0492}(1\sigma)_{-0.0751}^{+0.0711}(2\sigma)$ \ \ & \ \ $0.7665_{-0.1427}^{+0.1358}(1\sigma)_{-0.2069}^{+0.1835}(2\sigma)$\ \ \\
$\alpha$              \ \ & \ \ $-0.0256_{-0.1326}^{+0.1760}(1\sigma)_{-0.1907}^{+0.2730}(2\sigma)$ \ \ & \ \ $0.0198_{-0.1348}^{+0.1694}(1\sigma)_{-0.1960}^{+0.2606}(2\sigma)$\ \  & \ \ $0.0184_{-0.2922}^{+0.5293}(1\sigma)_{-0.3929}^{+0.9024}(2\sigma)$\ \ \\
$\Omega_{\Lambda}$    \ \ & \ \ $0.7371_{-0.0155}^{+0.0153}(1\sigma)_{-0.0236}^{+0.0223}(2\sigma)$\ \   & \ \ $0.7425_{-0.0146}^{+0.0137}(1\sigma)_{-0.0216}^{+0.0200}(2\sigma)$\ \  & \ \ $0.7425_{-0.0415}^{+0.0360}(1\sigma)_{-0.0617}^{+0.0497}(2\sigma)$\ \ \\
Age/Gyr               \ \ & \ \ $13.79_{-0.09}^{+0.09}(1\sigma)_{-0.14}^{+0.13}(2\sigma)$\ \            & \ \ $13.77_{-0.09}^{+0.09}(1\sigma)     _{-0.13}^{+0.13}(2\sigma)$\ \      & \ \ $13.77_{-0.13} ^{+0.17}(1\sigma)    _{-0.20}^{+0.26}(2\sigma)$\ \ \\
$\Omega_{\rm{m}}$     \ \ & \ \ $0.2629_{-0.0153}^{+0.0155}(1\sigma)_{-0.0223}^{+0.0236}(2\sigma)$\ \   & \ \ $0.2575_{-0.0146}^{+0.0137}(1\sigma)  _{-0.0200}^{+0.0216}(2\sigma)$\ \    & \ \ $0.2575_{-0.0360}^{+0.0415}(1\sigma)_{-0.0497}^{+0.0617}(2\sigma)$\ \ \\
$H_0$                 \ \ & \ \ $69.56_{-2.01}^{+2.14}(1\sigma)_{-2.91}^{+3.19}(2\sigma)$\ \            & \ \ $70.29_{-1.98}^{+1.95}(1\sigma)     _{-2.84}^{+2.91}(2\sigma)$\ \      & \ \ $70.29_{-5.08}  ^{+5.47}(1\sigma) _{-7.15}  ^{+7.87}(2\sigma)$\ \ \\
 \hline\hline
$\chi_{\rm min}^2$              \ \ & \ \ $502.266$\ \                  & \ \ $544.828$\ \                   & \ \ $ 36.930$\ \      \\
\hline\hline

 \end{tabular}
 \end{center}
 \caption{The best-fit values of  parameters $\Omega_{\rm{b}}h^2$, $A_S$, $\alpha$,
 $\Omega_{\Lambda}$, Age/Gyr, $\Omega_{m}$, and $H_0$ for the GCG model
 with the 1-$\sigma$ and 2-$\sigma$ uncertainties, as well as $\chi_{\rm min}^2$, 
for the data sets  SNe+CMB+BAO, SNe+GRBs+CMB+BAO, and GRBs+CMB+BAO,
respectively. }
 \end{table*}

\begin{figure}[!htbp]
\includegraphics[width=9.5cm]{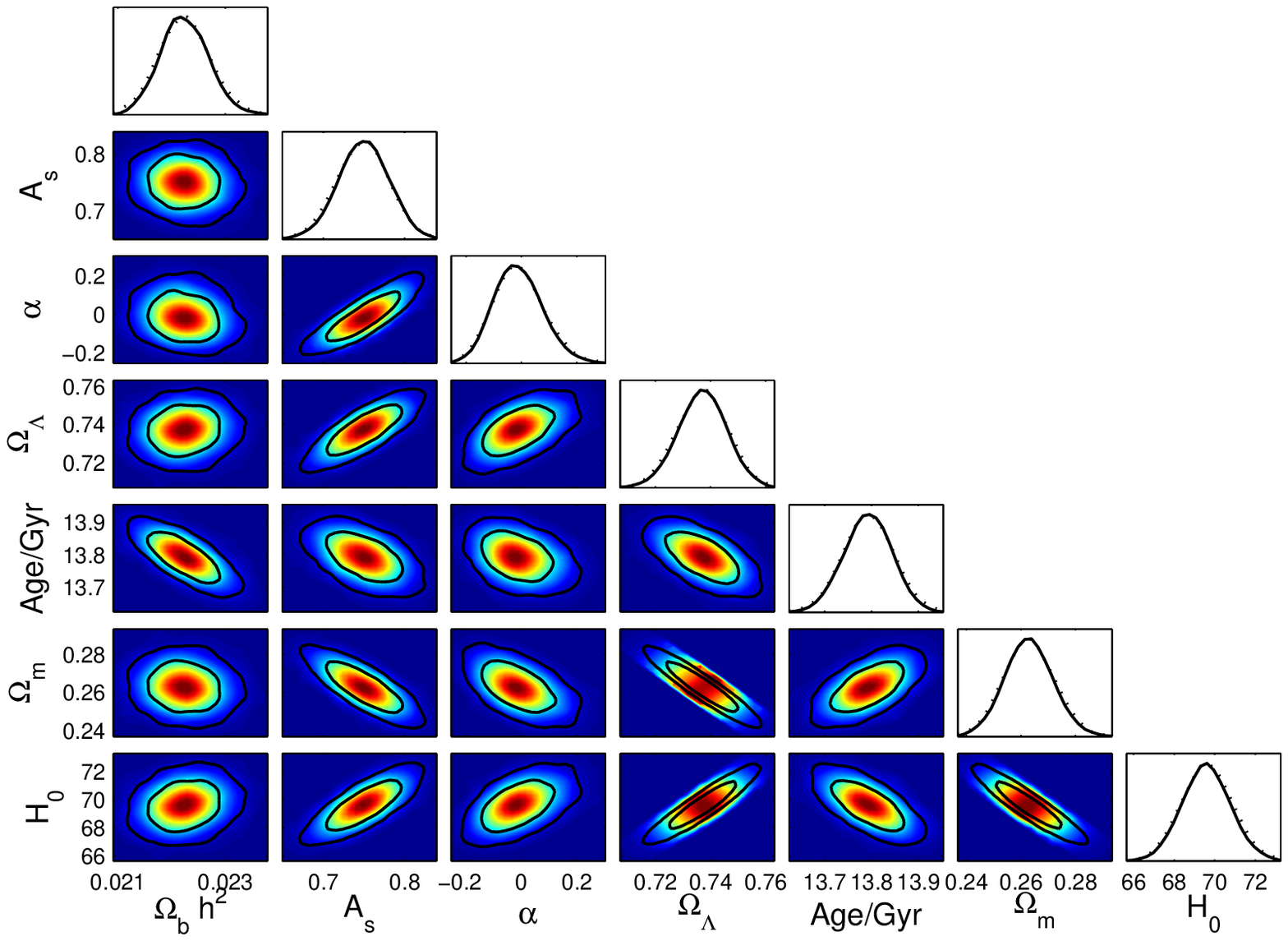}
\caption{The 2-D regions and 1-D marginalized distribution with the
1-$\sigma$ and 2-$\sigma$ contours of parameters
$\Omega_{\rm{b}}h^2$, $A_S$, $\alpha$, $\Omega_{\Lambda}$, Age/Gyr,
$\Omega_{\rm m}$, and $H_0$  in GCG model, for the data sets
SNe+GRBs+CMB+BAO.}
\end{figure}

\begin{figure}[!htbp]
\includegraphics[width=9.5cm]{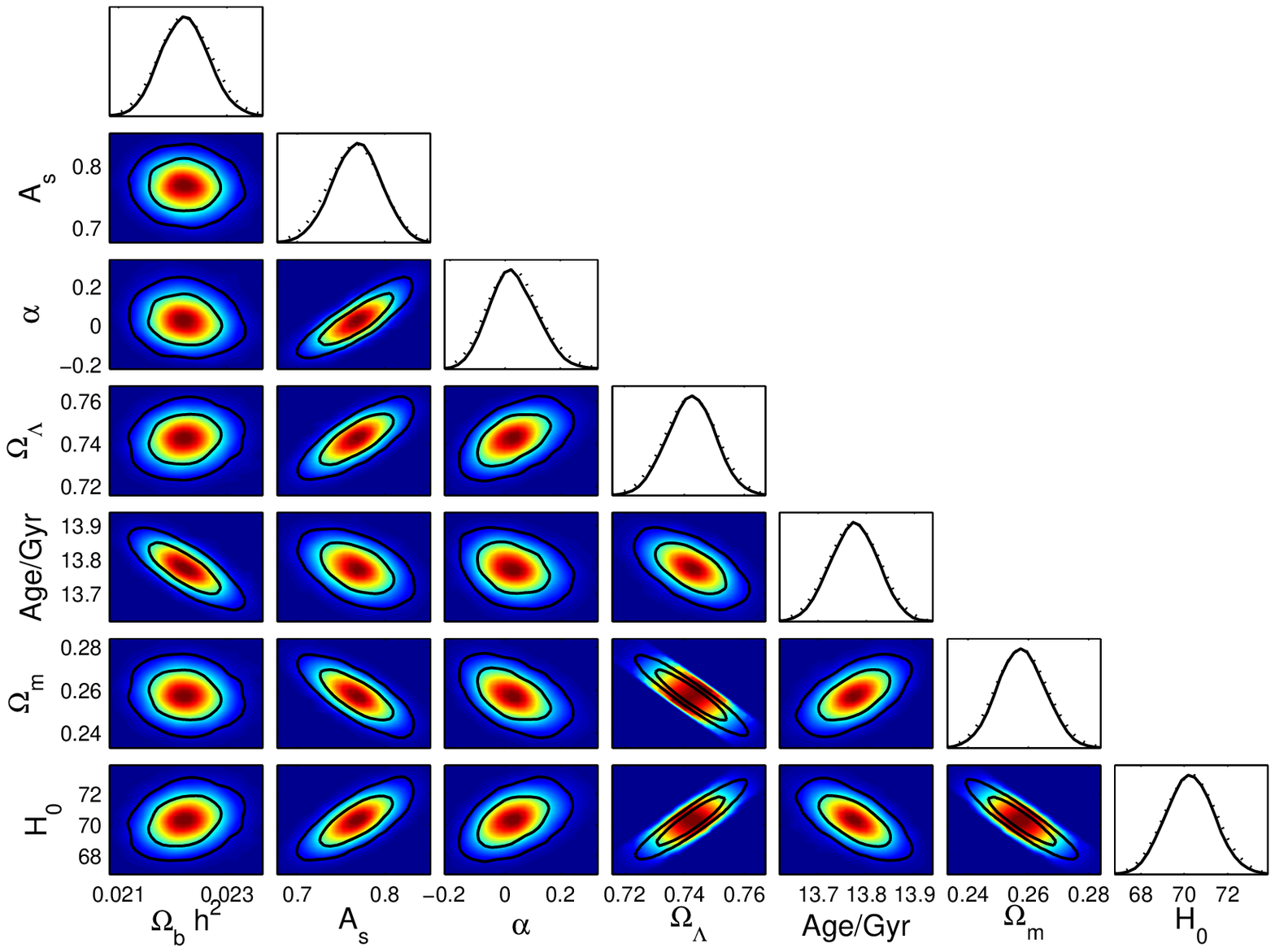}
\caption{The 2-D regions and 1-D marginalized distribution with the
1-$\sigma$ and 2-$\sigma$ contours of parameters
$\Omega_{\rm{b}}h^2$, $A_S$, $\alpha$, $\Omega_{\Lambda}$, Age/Gyr,
$\Omega_{\rm m}$, and $H_0$  in GCG model, for the data sets
SNe+CMB+BAO. }
\end{figure}

\begin{figure}[!htbp]
\includegraphics[width=9.5cm]{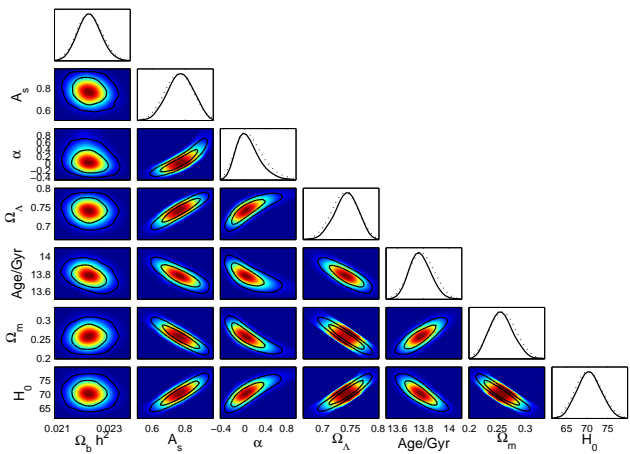}
\caption{The 2-D regions and 1-D marginalized distribution with the
1-$\sigma$ and 2-$\sigma$ contours of parameters
$\Omega_{\rm{b}}h^2$, $A_S$, $\alpha$, $\Omega_{\Lambda}$, Age/Gyr,
$\Omega_{\rm m}$, and $H_0$  in GCG model, for the data sets
GRBs+CMB+BAO.}
\end{figure}

\section{CONCLUSIONS}
By using the Markov Chain Monte Carlo method, we have constrained on
the generalized Chaplygin gas (GCG) model with the
cosmology-independent GRBs, as well as the Union2 SNe Ia set, the
CMB observation from WMAP7 result, and the BAO observation from SDSS
DR7 galaxy sample. With the joint observational data, the best-fit
values of the GCG model parameters  are
$A_S$=$0.7475_{-0.0539}^{+0.0556}(1\sigma)_{-0.0816}^{+0.0794}(2\sigma)$,
$\alpha$=$-0.0256_{-0.1326}^{+0.1760}(1\sigma)_{-0.1907}^{+0.2730}(2\sigma)$,
and the effective matter density $\Omega_{\rm
m}=0.2629_{-0.0153}^{+0.0155}(1\sigma)_{-0.0223}^{+0.0236}(2\sigma)$,
which are more stringent than  previous results for constraining the
GCG model parameters obtained using data of GRBs and/or other
combinations of observations.

\section*{Acknowledgements}
We thank Yun Chen,  Shuo Cao, Hao Wang,  Yan Dai, Chunhua Mao, Fang
Huang, Yu Pan, Jing Ming, Kai Liao  and Dr. Yi Zhang for
discussions. This work was supported by the National Science
Foundation of China under the Distinguished Young Scholar Grant
10825313, the Key Project Grants 10533010, and  by the Ministry of
Science and Technology national basic science Program (Project 973)
under grant No. 2007CB815401. LX acknowledges partial supports by
NSF (10703001), SRFDP (20070141034) of P.R. China and the
Fundamental Research Funds for the Central Universities (DUT10LK31).

\end{document}